\documentclass[aoas,MSNbibl,nameyear,dvips]{arximspdf}
\usepackage{graphicx}

\doi{10.1214/10-AOAS439}
\volume{5}
\issue{2B}
\pubyear{2011}
\firstpage{1657}
\lastpage{1677}

\makeatletter
\renewcommand{\cite}{\citet}

\newcommand{\dd}{\mathrm{d}}
\newcommand{\eqnnumber}{equation}
\newcommand{\eg}{e.g.}
\newcommand{\ie}{i.e.}
\newcommand{\Hipparcos}{\textit{Hipparcos}}
\newcommand{\normal}{{\mathcal N}}
\newcommand{\wishart}{{\mathcal W}}
\newcommand{\dirichlet}{{\mathcal D}}

\newcommand{\mm}{\mathbf{m}}
\newcommand{\bij}{\mathbf{b}_{ij}}
\newcommand{\mmj}{\mm_j}
\newcommand{\mmk}{\mm_k}
\newcommand{\vvi}{\mathbf{v}_i}
\newcommand{\wwi}{\mathbf{w}_i}
\newcommand{\ten}[1]{\mathbf{#1}} 
\newcommand{\BB}{\mathbf{B}}
\newcommand{\RR}{\mathbf{R}}
\renewcommand{\SS}{\mathbf{S}}
\newcommand{\TT}{\mathbf{T}}
\newcommand{\VV}{\mathbf{V}}
\newcommand{\PP}{\mbox{\bf P}}
\newcommand{\WW}{\mathbf{W}}
\newcommand{\II}{\mathbf{I}}
\newcommand{\BBij}{\BB_{ij}}
\newcommand{\RRi}{\RR_i}
\newcommand{\SSi}{\SS_i}
\newcommand{\VVj}{\VV_{\!j}} 
\newcommand{\TTij}{\TT_{ij}}
\newcommand{\TTik}{\TT_{ik}}
\newcommand{\T}{^{\scriptscriptstyle\top}} 
\newcommand{\alphaj}{\alpha_j}
\newcommand{\qij}{q_{ij}}
\newcommand{\qqj}{q_j}
\newcommand{\trace}{\operatorname{Trace}}
\newcommand{\norm}{|\!|}
\newcommand{\EPhi}{\langle\Phi\rangle}
\newcommand{\matrixleft}{\left[}
\newcommand{\matrixright}{\right]}

\newcommand{\parallax}{\pi}
\newcommand{\vrr}{v_r}
\newcommand{\ra}{\alpha}
\newcommand{\dec}{\ensuremath{\delta}}
\newcommand{\pmra}{\ensuremath{\mu_{\ra}}}
\newcommand{\pmdec}{\ensuremath{\mu_{\dec}}}
\newcommand{\vx}{\ensuremath{v_x}}
\newcommand{\vy}{\ensuremath{v_y}}
\newcommand{\vz}{\ensuremath{v_z}}
\newcommand{\AAA}{\ten{A}}
\newcommand{\arcsecs}{\textnormal{arcsec}}
\newcommand{\ngp}{\mathrm{NGP}}
\newcommand{\rangp}{\ra_\ngp}
\newcommand{\decngp}{\dec_\ngp}
\newcommand{\degree}{^{\circ}}
\makeatother

\begin{document}
\begin{frontmatter}

\title{Extreme deconvolution: Inferring complete distribution
functions from noisy, heterogeneous and incomplete
observations}
\runtitle{Extreme deconvolution}

\begin{aug}
\author[A]{\fnms{Jo} \snm{Bovy}\corref{}\thanksref{t1}\ead[label=e1]{jo.bovy@nyu.edu}},
\author[A]{\fnms{David W.} \snm{Hogg}\thanksref{t1,t2}\ead[label=e2]{david.hogg@nyu.edu}}
\and
\author[B]{\fnms{Sam T.} \snm{Roweis}\thanksref{t3}\ead[label=e3]{roweis@cs.nyu.edu}}
\thankstext{t1}{Supported in part by NASA (Grant
NNX08AJ48G) and the NSF (Grant AST-0908357).}
\thankstext{t2}{During part of the period in which this research was
performed, D. W. Hogg was a research fellow of the Alexander von Humboldt
Foundation of Germany at the Max-Planck-Institut f\"{u}r Astronomie,
Heidelberg, Germany.}
\thankstext{t3}{Deceased.}
\runauthor{J. Bovy, D. W. Hogg and S. T. Roweis}
\affiliation{New York University}
\address[A]{J. Bovy\\
D. W. Hogg\\
Center for Cosmology\\
\quad and Particle Physics\\
Department of Physics\\
New York University\\
4 Washington Place\\
New York, New York 10003\\
USA\\
\printead{e1}\\
\hphantom{E-mail: }\printead*{e2}} 
\address[B]{S. T. Roweis\\
Courant Institute\\
\quad of Mathematical Sciences\\
New York University\\
251 Mercer Street\\
New York, New York 10012\\
USA\\
\printead{e3}}

\end{aug}

\received{\smonth{10} \syear{2009}}
\revised{\smonth{11} \syear{2010}}

%
\begin{abstract}
We generalize the well-known mixtures of Gaussians approach to density
estimation and the accompanying Expectation--Maximization technique for
finding the maximum likelihood parameters of the mixture to the case
where each data point carries an individual $d$-dimen\-sional
uncertainty covariance and has unique missing data properties. This
algorithm reconstructs the error-deconvolved or ``underlying''
distribution function common to all samples, even when the individual
data points are samples from different distributions, obtained by
convolving the underlying distribution with the heteroskedastic
uncertainty distribution of the data point and projecting out the
missing data directions. We show how this basic algorithm can be
extended with conjugate priors on all of the model parameters and a
``split-and-merge'' procedure designed to avoid local maxima of the
likelihood. We demonstrate the full method by applying it to the
problem of inferring the three-dimensional velocity distribution of
stars near the Sun from noisy two-dimensional, transverse velocity
measurements from the \textit{Hipparcos} satellite.
\end{abstract}

%

\end{frontmatter}

\section{Introduction}

Inferring a distribution function given a finite set of samples from
this distribution function and the related problem of finding clusters
and/or overdensities in the distribution is a problem of significant
general interest [e.g.,
\cite{McLachlan1988}; \cite{Rabiner1993}; \cite{1998AJ1152384D}; Helmi et al.~(\citeyear{1999Natur40253H}); \cite{1999MNRAS308731S}; \cite{2005ApJ629268H}].
Performing this inference given only a~noisy set of measurements is a~problem commonly encountered in many of the sciences [see examples in
\cite{Carroll06a}]. In many cases of interest, the noise properties of
the observations are different from one measurement to the next (\ie,
they are heteroskedastic), even though the uncertainties are well
characterized for each observation. This is, for example, often the
case in astronomy, where in many cases the dominant source of
uncertainty is due to well-characterized photon-counting statistics,
while spatial and temporal variations in the atmosphere cause the
uncertainties to significantly vary even for sources of the same
apparent brightness observed with the same telescope.

The description you are interested in as a scientist is \textit{not} the
observed distribution, what you really want is the description of the
distribution that you would have if you had good data, that is, data
with vanishingly small uncertainties and with all of the dimensions
measured. In the low signal-to-noise regime the data never have these
two properties such that the underlying, true distribution cannot be
found without taking the noise properties of the data into account. If
you want to know the underlying distribution, in order to compare your
model with the data, you need to convolve the model with the data
uncertainties, not deconvolve the data. When the given set of data has
heterogeneous noise properties, that is, when the uncertainty
convolution is different for each data point, each data point is a
sample of a~different distribution, that is, the distribution obtained
from convolving the true, underlying distribution with the noise of
that particular observation. Incomplete data poses a similar problem
when the part of the data that is missing is different for different
data points.

Most existing approaches to density estimation only apply in the high~%
sig\-nal-to-noise regime [e.g.,
\cite{McLachlan1988}; \cite{Silverman86a}; \cite{diebolt94a}], and most approaches to
density estimation from noisy samples are nonparametric techniques that
assume that the noise is homoskedastic [e.g.,
\cite{Stefanski90a}; \cite{Zhang90a}]. The case of heteroskedastic
uncertainties has only recently attracted attention [e.g.,
\cite{Delaigle08a}; \cite{Staudenmayer08a}], and all of the approaches that
have been developed so far are nonparametric. None of these approaches
can be used when only incomplete data are available, although
parametric techniques that properly account for incomplete, but
noiseless, data have been developed
[Ghahramani and
Jordan (\citeyear{Ghahramani1994b}, \citeyear{Ghahramani1994})].

In this paper we show that the frequently used Gaussian-mixture-model
approach to density estimation can be generalized in the presence of
noisy, heterogeneous and incomplete data. The likelihood of the model
for each data point is given by the model convolved with the (unique)
uncertainty distribution of that data point; the objective function is
obtained by simply multiplying these individual likelihoods together
for the various data points. Optimizing this objective function, one
obtains a maximum likelihood estimate of the distribution (more
specifically, of its parameters).

While optimization of this objective function can, in principle, be
performed by a generic optimizer, we develop an
Expectation--Maximization (EM) algorithm that optimizes the objective
function. This algorithm works in much the same way as the normal EM
algorithm for mixture-of-Gaussians density estimation, except that an
additional degree of incompleteness is given by the actual values of
the observables, since we only have access to noisy projections of
these; in the expectation step these actual values are estimated based
on the noisy and incomplete measured values and the current estimate of
the distribution function. In the limit in which the noise is absent
but the data are lower dimensional projections of the quantities of
interest, this algorithm reduces to the algorithm described in
Ghahramani and
Jordan (\citeyear{Ghahramani1994b}, \citeyear{Ghahramani1994}).

We also show how prior distributions for a Bayesian version of the
calculation reporting a MAP estimate can be naturally included in this
algorithm as well as how a split-and-merge procedure that heuristically
searches parameter space for better approximations to the global
maximum can also be incorporated in this approach. These priors and the
split-and-merge procedure can be important when applying the EM
algorithm developed here in situations with real data where the
likelihood surface can have a very complicated structure. We also
briefly discuss the practical issues having to do with model selection
in the mixture model approach.

An application to a real data set is given in Section
\ref{sec:hipparcos}, where we fit the distribution of stellar
velocities near the Sun. The observed velocities of stars that we use
for this purpose have all of the properties that the approach developed
in this paper handles correctly: The velocity measurements are noisy,
and since we only use observations of the velocity components in the
plane of the sky, the data are incomplete, and this incompleteness is
different for each velocity measurement, which covers the full sky.
Nevertheless, we are able to obtain good agreement with other fits of
the velocity distribution based on complete data.

The technique we describe below has many applications besides returning
a~maximum likelihood fit to the error-deconvolved distribution function
of a~data sample. For instance, when an estimate of the uncertainty in
the estimated parameters or distribution function is desired or when a
full Bayesian analysis of the mixture model preferred, the outcome of
the maximum likelihood technique developed here can be used as a seed
for Markov Chain Monte Carlo (MCMC) methods for finite mixture modeling
[e.g., \cite{diebolt94a}; \cite{Richardson97a}].

\section{Likelihood of a mixture of Gaussian distributions given a set
of heterogeneous, noisy samples}\label{sec:objective}

Our goal is to fit a model for the distribution of a $d$-dimensional
quantity $\mathbf{v}$ using a set of $N$ observational data points~$\wwi$.
Therefore, we need to write down the probability of the data under the
model for the distribution. The observations are assumed to be noisy
projections of the true values $\vvi$,
%
\begin{equation}\label{eq:obs}
\wwi= \RRi\vvi+ \mbox{noise} ,
\end{equation}
where the noise is drawn from a Gaussian with zero mean and known
covariance matrix $\SSi$. The case in which there is missing data
occurs when the projection matrix $\RRi$ is rank-deficient.
Alternatively, we can handle the missing data case by describing the
missing data as directions of the covariance matrix that have a
formally infinite eigenvalue. In practice, we use very large
eigenvalues in the noise-matrix. When the data has only a small degree
of incompleteness, that is, when each data point has only a small
number of unmeasured dimensions, this latter approach is often the best
choice, since one often has some idea about the unmeasured values. For
example, in the example given below of inferring the velocity
distribution of stars near the Sun, we know that the stars are moving
at velocities that do not exceed the speed of light, which is not very
helpful, but also that none of the velocities exceed the local Galactic
escape speed, since we can safely assume that all the stars are bound
to the Galaxy. However, in situations in which each data point has
observations of a dimensionality $\ll$ $d$ using the projections
matrices will greatly reduce the computational cost, since, as will
become clear below, the most computationally expensive operations all
take place in the lower dimensional space of the observations.

We will model the probability density $p(\mathbf{v})$ of the true values
$\mathbf{v}$
as a~mix\-ture of $K$ Gaussians,
%
\begin{equation}
p(\mathbf{v}) = \sum_{j=1}^K \alphaj\normal(\mathbf{v}|\mmj,\VVj) ,
\end{equation}
where the amplitudes $\alphaj$ sum to unity and the function
$\normal(\mathbf{v}|\mm,\VV)$ is the Gaussian probability density function
with mean $\mm$ and variance matrix $\VV$,
%
\begin{equation}
\normal(\mathbf{v}|\mm,\VV) = (2\pi)^{-d/2} \det(\VV)^{-1/2}
\exp\bigl[-\tfrac{1}{2}(\mathbf{v}-\mm)\T\VV^{-1}(\mathbf{v}-\mm)\bigr].
\end{equation}

For a given observation $\wwi$ the likelihood of the model parameters
$\theta=(\alphaj,\mmj,\VVj)$ given that observation and the noise
covariance $\SSi$, which we will write as $p(\wwi|\theta)$, can be
written as
\begin{eqnarray}
p(\wwi|\theta) \equiv p(\wwi|\RRi,\SSi,\theta) &=&\sum_j\int
_\mathbf{v}\mathrm{d}\mathbf{v}\, p(\wwi,\mathbf{v},j|\theta)\nonumber\\ [-8pt]\\ [-8pt]
&=& \sum_j\int_\mathbf{v}\mathrm{d}\mathbf{v}\, p(\wwi|\mathbf{v})
p(\mathbf{v}|j,\theta)p(j|\theta),\nonumber
\end{eqnarray}
where
%
\begin{eqnarray}
p(\wwi|\mathbf{v}) &=& \normal(\wwi|\RRi\mathbf{v},\SSi),\nonumber\\
p(\mathbf{v}|j,\theta) &=& \normal(\mathbf{v}|\mmj,\VVj),\\
p(j|\theta) &=& \alphaj.\nonumber
\end{eqnarray}
This likelihood works out to be itself a mixture of Gaussians
%
\begin{equation}
p(\wwi|\theta) = \sum_j \alphaj\normal(\wwi|\RRi\mmj,\TTij),
\end{equation}
where
%
\begin{equation}
\TTij= \RRi\VVj\RRi\T+ \SSi .
\end{equation}

The free parameters of this model can now be chosen such as to maximize
an explicit, justified, scalar objective function $\phi$, given here by
the logarithm (log) likelihood of the model given the data, that is,
%
\begin{equation}\label{eq:totallike}
\phi= \sum_i \ln p(\wwi|\theta) = \sum_i \ln\sum_{j=1}^K \alphaj
\normal(\wwi|\RRi\mmj,\TTij).
\end{equation}
This function can be optimized in several ways, one of which is to use~%
a~ge\-neric optimizer to increase the likelihood until it reaches a
maximum. This approach is complicated by parameter constraints (\eg,
the amplitudes~$\alphaj$ must all be nonnegative and add up to one,
the variance matrices must be positive definite and symmetric) and
multimodality of the likelihood surface. In what follows we will
describe a different approach that is natural in this setting: An EM
algorithm that iteratively and monotonically maximizes the likelihood,
while naturally respecting the restrictions on the parameters.

\section{Fitting mixtures with heterogeneous, noisy data using an EM algorithm}

To optimize the likelihood in \eqnnumber\ (\ref{eq:totallike}), we can
extend the standard EM approach to Gaussian mixture estimation. In the
case of complete and precise observations, the problem is framed as a
tractable missing-data problem by positing that the labels or indicator
variables $\qij$ indicating which Gaussian $j$ a data point $i$ was
drawn from are missing [\cite{Dempster1977}]. We extend this approach
by including the true values $\vvi$ as extra missing data. This is a
well-known approach for handling measurement uncertainty in latent
variable or random effects models [e.g., \cite{Schafer93a}; \cite{Schafer96a}].

We write down the ``full data'' log likelihood---the likelihood we
would write down if we had the true values $\vvi$ and the labels
$\qij$,
%
\begin{equation}\label{eq:incompletefulllike}
\Phi= \sum_i \sum_j \qij\ln\alpha_j \normal(\vvi|\mmj,\VVj) .
\end{equation}
We will now show how we can use the EM methodology to find
straightforward update steps that maximize the full data likelihood of
the model. In Appendix \ref{sec:convergenceproof} we prove that these
updates also maximize the likelihood of the model given the noisy
observations.

The E-step consists as usual of taking the expectation of the full data
likelihood with respect to the current model parameters $\theta$.
Writing out the full data log likelihood from \eqnnumber\
(\ref{eq:incompletefulllike}), we find
\begin{eqnarray}
\Phi &=& \sum_i \sum_j \qij\biggl[ \ln\alpha_j -\frac{d}{2} \ln(2
\pi)\nonumber\\ [-8pt]\\ [-8pt]
&&\hphantom{\sum_i \sum_j \qij\biggl[}{}- \frac{1}{2} \ln\det\VVj-\frac{1}{2} (\mathbf{v}_i - \mmj)\T\VVj^{-1}
(\mathbf{v}_i - \mmj) \biggr],\nonumber
\end{eqnarray}
which shows that in addition to the expectation of the indicator
variables $\qij$ for each component we also need the expectation of the
$\qij\vvi$ terms and the expectation of the $\qij\vvi\vvi\T$ terms
given the data, the current model estimate and the component $j$. The
expectation of the $\qij$ is equal to the posterior probability that a
data point $\wwi$ was drawn from the component $j$. The expectation of
the $\vvi$ and the $\vvi\vvi\T$ can be found as follows: Consider the
joint distribution for the true and observed velocities, denoted by the
expanded vector $[\vvi\T\wwi\T]\T$, given the
model estimate and the component $j$. From the description of the
problem, we can see that this vector is distributed normally with mean
%
\begin{equation}\label{eq:combinedmean}
\mm' = \matrixleft\matrix{
\mmj\cr
\RRi\mmj
}
\matrixright
\end{equation}
and covariance matrix
%
\begin{equation}\label{eq:combinedcovar}
\VV' = \matrixleft\matrix{
 \VVj& \VVj\RRi\T \cr
 \RRi\VVj& \TTij
}
\matrixright.
\end{equation}
The conditional distribution of the $\vvi$ given the data $\wwi$ is
normal with mean
%
\begin{equation}\label{eq:bbij}
\bij\equiv\mmj+ \VVj\RRi\T\TTij^{-1} (\wwi- \RRi\mmj)
\end{equation}
and covariance matrix
%
\begin{equation}\label{eq:BBij}
\BBij\equiv\VVj- \VVj\RRi\T\TTij^{-1} \RRi\VVj.
\end{equation}
Thus, we see that the expectation of $\vvi$ given the data $\wwi$, the
model estimate and the component $j$ is given by $\bij$, whereas the
expectation of the $\vvi\vvi\T$ given the same is given by $\BBij+
\bij\bij\T$.

Given this, the expectation of the full data log likelihood is given by\vspace*{1pt}
\begin{eqnarray}\label{eq:incompletefulldataloglike}
\hspace*{20pt}\EPhi&=& \sum_{i,j} \qij \biggl[ \ln\alpha_j -\frac{d}{2}
\ln(2 \pi)\nonumber\\ [-8pt]\\ [-8pt]
&&\hphantom{\sum_{i,j} \qij \biggl[}{} - \frac{1}{2} \trace \bigl[\ln\VVj+ \bigl(\BBij+ (\mmj- \bij)(\mmj- \bij)\T\bigr)
\VVj^{-1}\bigr]\biggr].\nonumber
\end{eqnarray}
Straightforward optimization of this with respect to the model
parameters gives the following algorithm:\vspace*{1pt}
\begin{eqnarray}\label{eq:updateEMincomplete}
\mbox{\textit{E-step}:}\quad
\qij&\leftarrow& \frac{\alpha_j \normal(\wwi|\RRi\mmj,\TTij)}{\sum_k \alpha_k \normal(\wwi|\RRi\mmk,\TTik)},\nonumber\\
\bij&\leftarrow& \mmj+ \VVj\RRi\T\TTij^{-1} (\wwi- \RRi\mmj),\nonumber\\
\BBij&\leftarrow& \VVj- \VVj\RRi\T\TTij^{-1} \RRi\VVj,
\end{eqnarray}
\begin{eqnarray}
\hspace*{50pt}\mbox{\textit{M-step}:}\quad
\alphaj&\leftarrow& \frac{1}{N}\sum_i \qij,\nonumber\\
\mmj&\leftarrow& \frac{1}{\qqj}\sum_i \qij\bij,\nonumber\\
\VVj&\leftarrow& \frac{1}{\qqj}\sum_i \qij
[(\mmj-\bij)(\mmj-\bij)\T+ \BBij] ,\nonumber
\end{eqnarray}
where $\qqj=\sum_i \qij$.

In Appendix \ref{sec:convergenceproof} we prove that this procedure for
maximizing the full data likelihood also monotonically increases the
likelihood of the data $\wwi$ given the model, as is the case for the
EM algorithm for noiseless and complete measurements
[\cite{Dempster1977}; \cite{Wu1983}].

\section{Extensions to the basic algorithm}

Singularities and local maxima are two problems that can severely limit
the generalization capabilities of the computed density estimates for
inferring the densities of unknown data points. These are commonly
encountered when using the EM algorithm to iteratively compute the
maximum likelihood estimates of Gaussian mixtures: Singularities arise
when the covariance in \eqnnumber\ (\ref{eq:updateEMincomplete}) becomes
singular; the EM updates might get stuck in a local maximum because of
the monotonic increase in likelihood ensured by the EM algorithm. The
latter can be avoided through the use of a stochastic EM procedure
[\cite{Broniatowksi83}; \cite{Celeux85a}; \cite{Celeux86b}] or through the split and
merge procedure described below.

\subsection{Bayesian-inspired regularization}

The problem of singular covariances can be mitigated through the use of
priors on the model parameters in a Bayesian setting
[\cite{Ormoneit1995}]. It should be emphasized here that this
calculation is only Bayesian in the sense of producing a maximum a
posteriori (MAP) point estimate rather than a maximum likelihood
estimate, and that this is no different than penalized maximum
likelihood. We briefly show here that this procedure can be applied
here as well.

The regularization scheme of \cite{Ormoneit1995} introduces conjugate
priors on the Gaussian mixtures parameters space $\theta= (\alphaj,
\mmj,\VVj)$ as penalty terms. These conjugate priors are the following:
A normal density $\normal(\mmj|\hat\mm,\eta^{-1}\VVj)$ for the
mean of
each Gaussian, a Wishart density $\wishart(\VVj^{-1}|\omega,\WW)$
[\cite{Gelman00a}],
%
\begin{equation}
\wishart(\VVj^{-1}|\omega,\WW) =
c(\omega,\WW)|\VVj^{-1}|^{\omega-(d+1)/2}\exp[-\trace
[\WW\VVj^{-1}]],
\end{equation}
with $c(\omega,\WW)$ a normalization constant, for the covariance of
each Gaussian, and a~Dirichlet density $\dirichlet(\alpha|\gamma)$,
given by
%
\begin{equation}
\dirichlet(\alpha|\gamma) = b \prod_j \alphaj^{\gamma_j-1} ,
\end{equation}
where $b$ is a normalizing factor, for the amplitudes $\{\alphaj\}$.
Optimizing the posterior distribution for the model parameters replaces
the M-step of \eqnnumber~(\ref{eq:updateEMincomplete}) with
\begin{eqnarray}\label{updateBayes}
\hspace*{20pt}\alphaj&\leftarrow&\frac{\sum_i \qij+\gamma_j-1}{N+\sum_k \gamma
_k - K},\qquad \mmj\leftarrow\frac{\sum_i \qij\bij+ \eta\hat{\mm}}{\qqj+
\eta},\nonumber\\
\VVj&\leftarrow&\biggl(\sum_i \qij
[(\mmj-\bij)(\mmj-\bij)\T+\BBij]\nonumber\\ [-8pt]\\ [-8pt]
&&\hspace*{34pt}+\eta(\mmj-\hat{\mm})(\mmj-\hat{\mm})\T+ 2\WW\biggr)\nonumber\\
&&\Big/\qqj+1+2\bigl(\omega-(d+1)/2\bigr).\nonumber
\end{eqnarray}
Hyperparameters can be set by leave-one-out cross-validation. Vague
priors on the amplitude and the means are obtained by setting
%
\begin{equation}
\gamma_j = 1\qquad \forall j,\qquad \omega= (d+1)/2,\qquad \eta= 0.
\end{equation}
Since we are only interested in the MAP estimate, propriety of the
resulting posterior is not an issue with the improper prior resulting
from this choice.

The label-switching problem in Bayesian mixtures [\cite{Jasra05a}] is
not an issue for the maximization of the posterior distribution here.

\subsection{Avoiding local maxima}

The split and merge algorithm starts from the basic EM algorithm, with
or without the Bayesian regularization of the variances, and jumps into
action after the EM algorithm has reached a~maximum, which more often
than not will only be a local maximum. At this point, three of the
Gaussians in the mixture are singled out and two of these Gaussians are
merged, while the third Gaussian is split into two Gaussians
[\cite{Naonori1998}]. An alternative, but similar, approach to local
maxima avoidance is given by the birth and death moves in reversible
jump MCMC [\cite{Richardson97a}] or variational approaches
[\cite{Ghahramani00variationalinference}; \cite{Beal03}] to mixture modeling.
These moves do not conserve the number of mixture components and are
therefore less suited for our fixed-$K$ approach to mixture modeling.

Full details of the split and merge algorithm are given in
Appendix \ref{sec:splitmergeappendix}.

\subsection{Setting the remaining free parameters}\label{sec:Kw}

No real world application of Gaussian mixture density estimation is
complete without a well-specified methodology for setting the number of
Gaussian components $K$ and any hyperparameters introduced in the
Bayesian regularization described above, the covariance regularization
$\WW$. If we further assume that $\WW= w\II$, then this covariance
regularization parameter basically sets the square of the smallest
scale of the distribution function on which we can reliably infer
small-scale features. Therefore, this scale could be set by hand to the
smallest scale we believe we have access to based on the properties of
the data set.\looseness=-1

\begin{figure}

\includegraphics{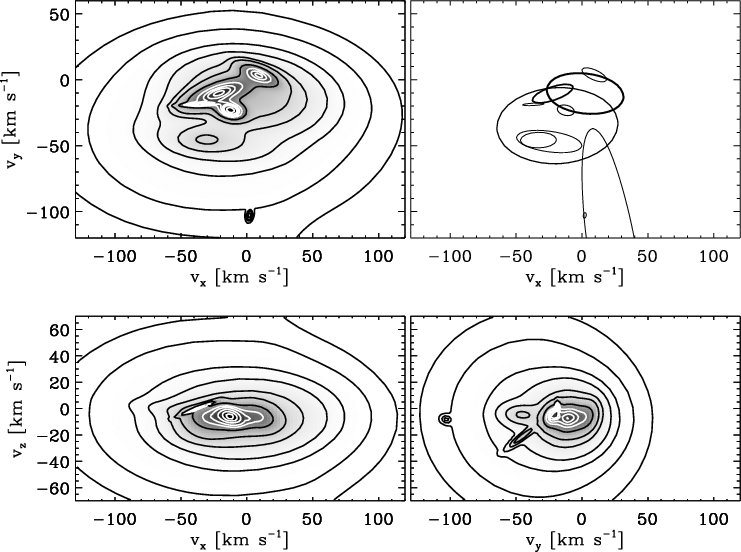}%
\vspace*{-3pt}
  \caption{Two-dimensional projections of the three-dimensional velocity
distribution of \Hipparcos\ stars using 10 Gaussians and $w = 4$
km$^2$ $\cdot$ s$^{-2}$. The top right plot shows 1-sigma covariance ellipses
around each individual Gaussian in the $v_x$--$v_y$ plane; the
thickness of each covariance ellipse is proportional to the natural
logarithm of its amplitude $\alpha_j$. In the other three panels the
density grayscale is linear and contours contain, from the inside
outward, 2, 6, 12, 21, 33, 50, 68, 80, 90, 95, 99 and 99.9 percent of
the distribution. 50~percent of the distribution is contained within
the innermost dark contour. The feature at $v_y \approx -100$ km $\cdot$ s$^{-1}$ is
real and corresponds to a known feature in the velocity
distribution: the Arcturus moving group; Indeed, all the features that
appear in these projections are real and correspond to known
features.}\label{fig:4veldist}
\vspace*{-3pt}
\end{figure}

In order to get the best results, the parameters $K$ and $w$ should be
set by some objective procedure. As mentioned above, leave-one-out
cross-validation [Stone~(\citeyear{stone74a})] could be used to set the
regularization parameter $w$, and the number of Gaussians could be set
by this procedure as well. Other techniques include methods based on
Bayesian model selection [\cite{roberts98a}] as well as approaches
based on minimum encoding inference
[\cite{wallace68a}; \cite{Oliver96a}; \cite{rissanen78a}; \cite{Schwartz78a}], although these
methods have difficulty dealing with significant overlap between
components (such as the overlap we see in the example in Figure
\ref{fig:4veldist}), but there are methods to deal with these
situations [\cite{baxter95a}]. Alternatively, when a separate, external
data set is available, we can use this as a test data set to validate
the obtained distribution function. All of these methods are explored
in an accompanying paper on the velocity distribution of stars in the
Solar neighborhood from measurements from the \Hipparcos\ satellite
[see below; \cite{Bovy09b}].

A rather different approach to the model selection problem is to avoid
it altogether. That is, by introducing priors over the hyperparameters
and including them as part of the model it is often possible to infer,
or fully marginalize over, them simultaneously with the parameters of
the components of the mixture. These methods also address uncertainty
quantification throughout the model. Such approaches include reversible
jump MCMC methods [\cite{Richardson97a}], mixtures consisting of an
infinite number of components based on the Dirichlet process
[\cite{rasmussen00infinite}], or approximate, variational algorithms
[\cite{Ghahramani00variationalinference}; \cite{Beal03}]. Extending these
approaches to deal with noisy, heterogeneous and incomplete data is
beyond the scope of this paper, but it is clear that this extension is,
in principle, straightforward: the MCMC methods mentioned above can
include the true values of the observations $\vvi$---known in the
Bayesian MCMC literature as data augmentation---and these can be Gibbs
sampled given the current model and the observed values $\wwi$ in an
MCMC sweep from the Gaussian with mean given in
\eqnnumber\ (\ref{eq:bbij}) and variance given in
\eqnnumber\ (\ref{eq:BBij}).

\section{The velocity distribution from \Hipparcos\ data}\label{sec:hipparcos}

We have applied the technique developed in this paper to the problem of
inferring the velocity distribution of stars in the Solar neighborhood
from transverse angular data from the \Hipparcos\ satellite and we
present in this section some results of this study to demonstrate the
performance of the algorithm on a real data set. A more detailed and
complete account of this study is presented elsewhere [\cite{Bovy09b}].

The local velocity distribution is interesting because we can learn
about the structure and evolution of the Galactic disk from deviations
from the smooth, close to Gaussian velocity distribution expected in
simple, axisymmetric models of the disk. It has been shown that the
dynamical effects of a~bar-shaped distribution of stars in the central
region of our Galaxy can resonate in the outer parts of the disk and
give rise to structure in the velocity distribution [e.g.,
\cite{Dehnen00a}; \cite{Bovy10b}]. Similarly, steady-state or transient spiral
structure can effect the velocities of stars in a coherent way, such
that we can see this effect locally [e.g.,
\cite{2005AJ130576Q}; \cite{2004MNRAS350627D}]. Inferring the local
velocity distribution from observational data is therefore necessary to
assess whether these dynamical signatures are observed.

Velocities of stars are not directly observable. Rather, they need to
be patched together from observations of the stars' directions on the
sky at different times---the branch of astronomy known as
astrometry---and spectroscopic observations to determine the velocity
along the line of sight. The annual motion of the Earth around the Sun
gives rise to an apparent displacement of a star relative to background
objects that is inversely proportional to the distance to the star.
Measurements of this apparent shift, or parallax, can thus be used to
determine the distance to stars. Parallaxes are traditionally reported
in units of arcseconds; a star with a parallax of 1 arcsecond is
defined to be at a distance of 1 parsec (pc), which equals $3\times
10^{16}$~m. The intrinsic motion of a star also gives rise to a
systematic shift in its position relative to background sources, such
that its angular motion---known as its proper motion---can be measured.
Combining the distance and angular velocity gives the components of the
space velocity of a star that are perpendicular to the line of sight.

The astrometric ESA space mission \Hipparcos, which collected data over
a~3.2~year period around 1990, provided for the first time an all-sky
catalogue of absolute parallaxes and proper motions, with typical
uncertainties in these quantities on the order of milli-arcseconds
[\cite{ESA97a}]. From this catalogue of $\sim$100,000 stars,
kinematically unbiased samples of stars with accurate positions and
velocities can be extracted [\cite{1998MNRAS298387D}]. Since this
was a purely astrometric mission, and the only components of a~star's
velocity that can be measured astrometrically are the components
perpendicular to the line of sight, the line-of-sight velocities of the
stars in the \Hipparcos\ sample were not obtained during the mission.

Distances in astronomy are notoriously hard to measure precisely, and
at the accuracy level of the \Hipparcos\ mission distances can only be
reliably obtained for stars near the Sun (out to $\sim$100 pc; the
diameter of the Galactic disk is about 30,000 pc). In addition to this,
since distances are measured as inverse distances (parallaxes), only
distances that are measured relatively precisely will have
approximately Gaussian uncertainties associated with them. Balancing
the size of the sample with the accuracy of the distance measurement
leaves us with distance uncertainties that are typically
$\sim$10-percent, such that the velocities perpendicular to the line
of sight that are obtained from the proper motions and the distances
have low signal-to-noise. Since the dominant source of noise is due to
simple photon-counting statistics [\cite{2007ASSL250V}], the
uncertainties are well characterized and can be assumed to be known, as
is necessary for the technique developed in this paper to apply.
Star-to-star correlations are negligible and can be ignored
[\cite{2007ASSL250V}].

Of course, if we want to describe the distribution of the velocities of
the stars in this sample, we need to express the velocities in a common
reference frame, which for kinematical studies of stars around the Sun
is generally chosen to be the Galactic coordinate system, in which the
$x$-axis points toward the Galactic center, the $y$-axis points in the
direction of Galactic rotation, and the $z$-axis points toward the
North Galactic Pole [\cite{1960MNRAS121123B}; \cite{1998gaasbookB}].
The measured velocities perpendicular to the line of sight are then
projections of the three-dimensional velocity of a star with respect to
the Sun in the two-dimensional plane perpendicular to the line of sight
to the star. Therefore, this projection is different for each
individual star.

Observations of celestial objects are expressed in the equatorial
coordinate system, in which the Earth's geographic poles and equator
are projected onto the celestial sphere. The components of the
three-dimensional velocities $\mathbf{v}$ of the stars in the Galactic
coordinate system in terms of the observed quantities in the equatorial
coordinate frame---angular position on the sky ($\ra$, $\dec$), inverse
distance ($\parallax$), angular motion on the sky ($\pmra$, $\pmdec$), and
line-of-sight velocity ($\vrr$)---are given by
%
\begin{equation}\label{eq:vrpmrapmdectoUVW2}
\mathbf{v}\equiv\matrixleft\matrix{
 \vx\cr \vy\cr \vz
}
\matrixright= \TT \AAA \matrixleft\matrix{
 \vrr \cr
\dfrac{k}{\parallax}\pmra\cos\dec\cr
\dfrac{k}{\parallax}\pmdec
}
\matrixright,
\end{equation}
where $k = 4.74047$, $[\vrr] = \mathrm{km}\cdot \mathrm{s}^{-1}$, $[\parallax] = \arcsecs,
[\pmra]=[\pmdec]= \arcsecs\ \cdot \mathrm{yr}^{-1}$. The matrix $\TT$ transforms the
velocities from the equatorial reference frame in which the
observations are made to the Galactic coordinate frame; it depends on a
few parameters defining this coordinate transformation and is given by
%
\begin{eqnarray}\label{eq:radectolbT}
\TT &=& \matrixleft\matrix{
\cos\theta& \sin\theta& 0\cr
\sin\theta& -\cos\theta& 0\cr
0&0&1
}
\matrixright
\matrixleft\matrix{
 -\sin\decngp& 0 & \cos\decngp\cr
  0 & 1 & 0 \cr
\cos\decngp& 0 & \sin\decngp
}
\matrixright\\
&&{} \times\matrixleft\matrix{
\cos\rangp
&\sin\rangp&0\cr
-\sin\rangp& \cos\rangp & 0\cr
 0&0&1
}
\matrixright.
\end{eqnarray}
The matrix $\TT$ depends on the epoch that the reduced data are
referred to (1991.25 for \Hipparcos) through the values of $\rangp$,
$\decngp$  and $\theta$ (the position in equatorial coordinates of the
north Galactic pole, and the Galactic longitude of the north Celestial
pole, respectively). These quantities were defined for the epoch 1950.0
as follows: [\cite{1960MNRAS121123B}]: $\rangp=
12\ \mathrm{h}\ 49\ \mathrm{m}$, $\decngp= 27\degree.4$, and
$\theta= 123\degree$.

The matrix $\AAA$ depends on the position of the source on the sky,
%
\begin{equation}
\AAA= \matrixleft\matrix{
 \cos\ra& -\sin\ra& 0 \cr
 \sin \ra& \cos\ra& 0 \cr
 0 & 0& 1
}
\matrixright\matrixleft\matrix{
\cos\dec& 0 & -\sin\dec\cr
0 & 1 & 0 \cr
\sin\dec
& 0 & \cos\dec
}
\matrixright .
\end{equation}
In the context of the deconvolution technique described above, the
observations are $\mathbf{w}\equiv\matrixleft\matrix{ \vrr, &
\frac{k}{\parallax}\pmra\cos\dec, & \frac{k}{\parallax}\pmdec}
\matrixright\T$ and the
projection matrix is $\RR^{-1} \equiv\TT\AAA$. Since we do not use
the radial velocities of the stars, we set~$\vrr$ to zero in $\mathbf
{w}$ and
use a large uncertainty-variance for this component of the uncertainty
variance $\SSi$; equivalently, we could remove $\vrr$ from $\mathbf{w}$ and
restrict~$\RR$ to the projection on the sky.

We have studied the velocity distribution of a sample of main sequence
stars selected to have accurate distance measurements (parallax
uncertainties $\sigma_\pi/\pi< 0.1$) and to be kinematically unbiased
(in that the sample of stars faithfully represents the kinematics of
similar stars). In detail, we use the sample of 11,865 stars from
\cite{1998MNRAS298387D}, but we use the new reduction of the
\Hipparcos\ raw data, which has improved the accuracy of the
astrometric quantities
[van Leeuwen~(\citeyear{2007ASSL250V}; \citeyear{2007AA474653V})]. A particular
reconstruction of the underlying velocity distribution of the stars\vadjust{\goodbreak} is
shown in Figure \ref{fig:4veldist}, in which 10 Gaussians are used, a
prior on the variances was used (the prior was restricted to $\WW=
w\II$), and this regularization parameter $w$ is set to 4 km$^2\cdot\mathrm{s}^{-2}$.

These values for the hyperparameters were set using an external data
set rather than any of the other methods described in
Section \ref{sec:Kw}. For this we use a set of 7682 stars from the
Geneva--Copenhagen Survey
[\cite{2004AA418989N}; \cite{2008arXiv08113982H}] for which the
line-of-sight velocity (perpendicular to the plane of the sky) has been
measured spectroscopically. This is a separate data set from the one
considered above. It partially overlaps with the previous data set and
it is also kinematically unbiased. We then fit the velocity
distribution for different choices of the hyperparameters and evaluate
the probability of the line-of-sight velocities for the best-fit
velocity distribution based on tangential velocities. The values of the
hyperparameters that maximize this probability are $K = 10$ and
$w={}$4~km$^2\cdot~\mathrm{s}^{-2}$.

The recovered distribution compares favorably with other
reconstructions of the velocity distribution of stars in the Solar
neighborhood, based on the same sample of stars [using a maximum
penalized likelihood density estimation technique,
\cite{1998AJ1152384D}], as well as with those based on other
samples of stars for which three-dimensional velocities are available
[\cite{1999MNRAS308731S}; \cite{2004AA418989N}; \cite{2005AA430165F}; \cite{2008AA490135A}].
In particular, this means that the main shape of the velocity
distribution agrees with that found in previous studies and that the
number and location of the peaks in the distribution, all real and
known features, are consistent with those found before. This includes
the very sparsely populated feature at $v_y \approx-100$ km${}\cdot{}$s$^{-1}$,
which is known as the Arcturus moving group. Therefore, we conclude
that the method developed in this paper performs very well on this
complicated data set. In contrast to previous determinations of the
velocity distributions, our method allows us to study the structures
found quantitatively, since it turns out that individual structures in
the velocity distribution are well represented by individual components
in the mixture model. Thus, we were able to conclude that these
structures are not the remnants of a large group of stars that formed
together in a cluster, but rather that they are probably caused by
dynamical effects related to the bar at the center of the Milky Way or
spiral structure [\cite{Bovy10a}].

\begin{figure}

\includegraphics{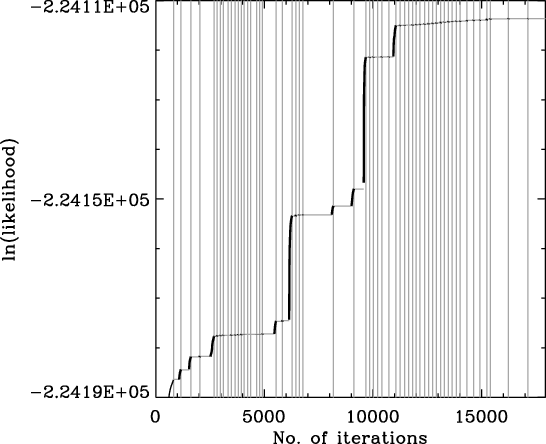}
\vspace*{-3pt}
  \caption{Convergence of the full algorithm: total log likelihood at
  each iteration step. Shown are only split-and-merge steps that
  improve the likelihood; each vertical gray line corresponds to a
  point at which a successful split and merge is performed. For
  clarity's sake, we show in black only the parts of the
  split-and-merge steps at which the likelihood is larger than the
  likelihood right before that split-and-merge procedure; the log
  likelihoods of the steps in a split-and-merge procedure in which the
  likelihood is still climbing back up to the previous maximum in
  likelihood have been replaced by horizontal gray segments. The
  $y$-axis has been cut off for display purposes: The log likelihood
  of the initial condition was $-$2.39E$-$5.}\label{fig:convergence}
  \vspace*{-3pt}
\end{figure}

The convergence of the algorithm is shown in Figure
\ref{fig:convergence}. Only split-and-merge steps that improved the
likelihood are shown in this plot, therefore, the actual number of
iterations is much higher than the number given on the $x$-axis. It is
clear that all of the split-and-merge steps only slightly improve the
initial estimate from the first EM procedure, but since what is shown
is the likelihood per data point, the improvement of the total
likelihood is more significant.

\setcounter{footnote}{3}
\section{Implementation and code availability}

The algorithm presented in this paper was implemented in the C
programming language, depending only on the standard C library and the
GNU Scientific
Library.\footnote{\url{http://www.gnu.org/software/gsl/}.} The code is
available at \href{http://www.code.google.com/p/extreme-deconvolution/}{http://code.google.com/p/extreme-deconvolution/};
instructions for its installation and use are given there. The code can
be compiled into a shared object library, which can then be
incorporated into other projects or accessed through
IDL\footnote{\url{http://www.ittvis.com/ProductServices/IDL.aspx}.} or
Python\footnote{\url{http://www.python.org/}.} wrapper functions
supplied with the~C~code.

The code can do everything described above. The convergence of a single
run is quick, but when including split-and-merge iterations the
convergence is rather slow because of the large number of
split-and-merge steps that can be taken by the algorithm (the
split-and-merge aspect of the algorithm, however, can easily be turned
off or restricted by setting the parameter specifying the number of
steps to go down the split-and-merge hierarchy).

\section{Conclusions and future work}

We have generalized the mixture of Gaussians approach to density
estimation such that it can be applied to noisy, heterogeneous and
incomplete data. The objective function is obtained by integrating over
the unknown true values of the quantities for which we only have noisy
and/or incomplete observations. In order to optimize the objective
function resulting from this marginalization, we have derived an EM
algorithm that monotonically increases the model likelihood; this EM
algorithm, in which the E-step involves finding the expected value of
the first and second moments of the true values of the observables
given the current model and the noisy observations, reduces to the
basic EM algorithm for Gaussian mixture modeling in the limit of
noiseless data. We have shown that the model can incorporate conjugate
priors on all of the model parameters without losing any of its
analytical attractiveness and that the algorithm can accommodate the
split-and-merge algorithm to deal with the presence of local maxima,
which this EM algorithm, as many other EM algorithms, suffers from.

The work presented here can be extended to be incorporated in various
more nonparametric approaches to density modeling, for example, in
mixture models with an infinite number of components based on the
Dirichlet Process [e.g., \cite{rasmussen00infinite}]. In this way
current advances in nonparametric modeling can be applied to the low
signal-to-noise sciences where the situation of complete and noise-free
data is more often than not an untenable and unattainable
approximation.

\begin{appendix}
\section{\texorpdfstring{Proof that the proposed algorithm maximizes the likelihood}%
{Appendix A: Proof that the proposed algorithm maximizes the likelihood}}\label{sec:convergenceproof}

We use Jensen's inequality in the continuous case for a concave
function~$f$ and a nonnegative integrable function $q$, where we have
assumed that $q$ is normalized, that is, $q$ is a probability
distribution. For each observation $\mathbf{w}$ we can then introduce a
function $q(\mathbf{v},j)$ such that
%
\begin{eqnarray}
\ln p(\mathbf{w}|\theta) &=& \ln\sum_j \int_{\mathbf{v}}
\mathrm{d}\mathbf{v}\,
p(\mathbf{w},\mathbf{v},j|\theta)\nonumber\\
&\geq& \sum_j
\int_{\mathbf{v}}\mathrm{d}\mathbf{v}\, q(\mathbf{v},j)\ln\frac{p(\mathbf
{w},\mathbf{v},j|\theta)}{q(\mathbf{v},j)} =
F(\mathbf{w}|q,\theta),\\
\ln p(\mathbf{w}|\theta) &\geq& F(\mathbf{w}|q,\theta) = \langle\ln
p(\mathbf{w},\mathbf{v},j|\theta) \rangle_{q} + {\mathcal H}(q),\nonumber
\end{eqnarray}
where ${\mathcal H}$ is the entropy of the distribution $q(\mathbf{v},j)$. This
inequality becomes an equality when we take
%
\begin{equation}
q(\mathbf{v},j) = p(\mathbf{v},j|\mathbf{w},\theta)  .
\end{equation}

The above holds for each data point, and we can write
%
\begin{equation}
p(\mathbf{v},j|\wwi,\theta) = p(\mathbf{v}|\wwi,\theta,j)  p(j|\wwi
,\theta)  .
\end{equation}
The last factor reduces to calculating the posterior probabilities
$\qij=p(j|\wwi,\break\theta)$ and we can write the $F$ function as (we drop
the entropy term here, since it plays no role in the optimization, as
it does not depend on the model parameters)
\begin{eqnarray*}
F &=& \sum_{i,j} \qij \int_{\mathbf{v}}\mathrm{d}\mathbf{v}\, p(\mathbf
{v}|\wwi,\theta,j) \ln p(\wwi,\mathbf{v},j|\theta)\\
&=& \sum_{i,j} \qij  \biggl[ \ln\alpha_j + \int_{\mathbf{v}}\mathrm
{d}\mathbf{v}\, p(\mathbf{v}|\wwi,\theta,j) \\
&& \hphantom{\sum_{i,j} \qij  \biggl[ \ln\alpha_j + \int_{\mathbf{v}}}{}\times\biggl[ -\frac{1}{2} \ln\det
\VV_j -\frac{1}{2} (\mathbf{v}-\mmj)\T\VVj^{-1}(\mathbf{v}-\mmj
)\biggr] \biggr].
\end{eqnarray*}
This shows that this reduces exactly to the procedure described above,
that is, to taking the expectation of the $\vvi$ and $\vvi\vvi\T$ terms
with respect to the distribution of the $\vvi$ given the data $\wwi$,
the current parameter estimate and the component $j$. We conclude that
the E-step as described above ensures that the expectation of the full
data log likelihood becomes equal to the log likelihood of the model
given the observed data. Optimizing this log likelihood in the M-step
then also increases the log likelihood of the model given the
observations. Therefore, the EM algorithm we described will increase
the likelihood of the model in every iteration, and the algorithm will
approach local maxima of the likelihood. Convergence is identified, as
usual, as extremely small incremental improvement in the log likelihood
per iteration.

\section{\texorpdfstring{Split and merge algorithm}{Appendix B: Split and merge algorithm}}\label{sec:splitmergeappendix}

Let us denote the indices of the three selected Gaussians as $j_1, j_2$
and $j_3$, where the former two are to be merged while $j_3$ will be
split. The Gaussians corresponding to the indices $j_1$ and $j_2$ will
be merged as follows: the model parameters of the merged Gaussian
$j_1'$ are
\begin{eqnarray}\label{merged}
\alpha_{j_1'} &=& \alpha_{j_1} + \alpha_{j_2},\nonumber\\ [-8pt]\\ [-8pt]
\theta_{j_1'}
&=& \frac{\theta_{j_1}q_{j_1} + \theta_{j_2}q_{j_2}}{q_{j_1} +
q_{j_2}},\nonumber
\end{eqnarray}
where $\theta_j$ stands for $\mmj$ and $\VVj$. Thus, the mean and the
variance of the new Gaussian is a weighted average of the means and
variances of the two merging Gaussians.

The Gaussian corresponding to $j_3$ is split as follows:
\begin{eqnarray}\label{split}
\alpha_{j_2'} &=& \alpha_{j_3'}= \alpha_{j_3}/2,\nonumber\\ [-8pt]\\ [-8pt]
\VV_{j_2'}&=&\VV_{j_3'} = \det(\VV_{j_3})^{1/d}\II.\nonumber
\end{eqnarray}
Thus, the Gaussian $j_3$ is split into equally contributing Gaussians
with each new Gaussian having a covariance matrix that has the same
volume as $\VV_{j_3}$. The means $\mm_{j_2'}$ and $\mm_{j_3'}$ can be
initialized by adding a random perturbation vector $\varepsilon_{j_m}$ to
$\mm_{j_3}$, for example,
%
\begin{equation}\label{split2}
\mm_{j_m'} = \mm_{j_3} + \varepsilon_{j_m} ,
\end{equation}
where $\norm\varepsilon_{j_m}\norm^2 \ll\det(\VV_{j_3})^{1/d}$ and
$m=1,2$.

After this split and merge initialization, the parameters of the three
affected Gaussians need to be re-optimized in a model in which the
parameters of the unaffected Gaussians are held fixed. This can be done
by using the M-step in \eqnnumber~(\ref{eq:updateEMincomplete}) for
the parameters of the three affected Gaussians, while keeping the
parameters of the other Gaussians fixed, including the amplitudes. This
ensures that the sum of the amplitudes of the three affected Gaussians
remains fixed. This procedure is called the \textit{partial EM procedure}.
After convergence this is then followed by the full EM algorithm on the
resulting model parameters. Finally, the resulting parameters are
accepted if the total log likelihood of this model is greater than the
log likelihood before the split and merge step. If the likelihood
does not increase, the same split and merge procedure is performed on
the next triplet of split and merge candidates.

The question that remains to be answered is how to choose the 2
Gaussians that should be merged and the Gaussian that should be split.
In general, there are $K(K-1)(K-2)/2$ possible triplets like this which
quickly reach a large number when the number of Gaussians $K$ gets
larger. In order to rank these triplets, one can define a \textit{merge
criterion} and a \textit{split criterion}.

The merge criterion is constructed based on the observation that if
many data points have equal posterior probabilities for two Gaussians,
these Gaussians are good candidates to be merged. Therefore, one can
define the merge criterion:
%
\begin{equation}
J_{\mathrm{merge}}(j,k|\theta) = \PP_j(\theta)\T\PP_k(\theta) ,
\end{equation}
where $\PP_j(\theta)=(q_{i1},\ldots,q_{iN})\T$ is the $N$-dimensional
vector of posterior probabilities for the $j$th Gaussian. Pairs of
Gaussians with larger $J_{\mathrm{merge}}$ are good candidates for a
merger.

We can define a split criterion based on the Kullback--Leibler distance
between the local data density around the $l$th Gaussian, which can be
written in the case of complete data as $p_l(\mathbf{w}) = 1/q_l\sum_i
q_{il}\delta(\mathbf{w}-\wwi)$, and the $l$th Gaussian density
specified by
the current model estimates $\mm_l$ and $\VV_l$. The Kullback--Leibler
divergence between two distributions $p(x)$ and $q(x)$ is given by
[\cite{Mackay2003}]
%
\begin{equation}
D_{\mathrm{KL}}(P\parallel Q) = \int\dd x\,p(x)\ln\frac{p(x)}{q(x)} .
\end{equation}
Since the local data density is only nonzero at a finite number of
values, we can write this as
%
\begin{equation}
J_{\mathrm{split}}(l|\theta) = \frac{1}{q_l}\sum_i q_{il} \biggl[\ln
\biggl(\frac{q_{il}}{q_l}\biggr) - \ln\normal(\wwi|\mm_l,\VV_l)
\biggr] .
\end{equation}
The larger the distance between the local density and the Gaussian~%
represen\-ting it, the larger $J_{\mathrm{split}}$ and the better candidate
this Gaussian is to be split.

When dealing with incomplete data determining the local data density is
more problematic. One possible way to estimate how well a particular
Gaussian describes the local data density is to calculate the
Kullback--Leibler divergence between the model Gaussian under
consideration and each individual data point perpendicular to the
unobserved directions for that data point. Thus, we can write
%
\begin{equation}
J_{\mathrm{split}}(l|\theta) = \frac{1}{q_l}\sum_i q_{il} \biggl[\ln
\biggl(\frac{q_{il}}{q_l}\biggr) - \ln
\normal(\wwi|\RRi\mm_l,\RRi\VV_l\RRi\T)\biggr] .
\end{equation}

Candidates for merging and splitting are then ranked as follows: first
the merge criterion $J_{\mathrm{merge}}(j,k|\theta)$ is calculated for
all pairs $j,k$ and the pairs are ranked by decreasing
$J_{\mathrm{merge}}(j,k|\theta)$. For each pair in this ranking the
remaining Gaussians are then ranked by decreasing
$J_{\mathrm{split}}(l|\theta)$.

To summarize the full algorithm, we briefly list all the steps
involved:
\begin{enumerate}
\item Run the EM algorithm as specified in \eqnnumber s
(\ref{eq:updateEMincomplete}) and (\ref{updateBayes}). Store the
resulting model parameters $\theta^*$ and the corresponding model log
likelihood~$\phi^*$.
\item Compute the merge criterion
$J_{\mathrm{merge}}(j,k|\theta^*)$ for all pairs $j,k$ and the split
criterion $J_{\mathrm{split}}(l|\theta^*)$ for all $l$. Sort the split
and merge candidates based on these criteria as detailed above.
\item
For the first triplet $(j,k,l)$ in this sorted list set the initial
parameters of the merged Gaussian using \eqnnumber\ (\ref{merged}) and
the parameters of the two Gaussian resulting from splitting the third
Gaussian using \eqnnumber s (\ref{split}) and (\ref{split2}). Then run the
partial EM procedure on the parameters of the three affected Gaussians,
that is, run EM while keeping the parameters of the unaffected
Gaussians fixed, and follow this up by running the full EM procedure on
all the Gaussians. If after convergence the new log likeli\-hood~$\phi$
is greater than $\phi^*$, accept the new parameter values $\theta^*
\leftarrow\theta$ and return to step two. If $\phi< \phi^*$, return
to the beginning of this step and use the next triplet $(j,k,l)$ in the
list.
\item Halt this procedure when none of the split and merge
candidates improve the log likelihood or, if this list is too long, if
none of the first $C$ lead to an improvement.
\end{enumerate}

Deciding when to stop going down the split-and-merge hierarchy will be
dictated in any individual application of this technique by
computational constraints. This is an essential feature of any
search-based approach to finding global maxima of (likelihood)
functions.

\end{appendix}

\section*{Acknowledgments}
It is a pleasure to thank Fr{\'{e}}d{\'{e}}ric Arenou and Phil Marshall
for comments and assistance and the anonymous referee and Associate
Editor for valuable criticism.


\printaddresses


\begin{thebibliography}{54}

%
%
\bibitem[\protect\citeauthoryear{{Antoja} et al.}{2008}]{2008AA490135A}
\begin{barticle}[author]
\bauthor{\bsnm{{Antoja}},~\bfnm{T.}\binits{T.}},
\bauthor{\bsnm{{Figueras}},~\bfnm{F.}\binits{F.}},
\bauthor{\bsnm{{Fern{\'{a}}ndez}},~\bfnm{D.}\binits{D.}} \AND
\bauthor{\bsnm{{Torra}},~\bfnm{J.}\binits{J.}}
(\byear{2008}). \btitle{{Origin and evolution of moving groups. I.
Characterization in the
observational kinematic-age-metallicity space}}.
\bjournal{Astron.  Astrophys.}
\bvolume{490}
\bpages{135}.
\end{barticle}
\endbibitem

%
%
\bibitem[\protect\citeauthoryear{{Baxter}}{1995}]{baxter95a}
\begin{btechreport}[author]
\bauthor{\bsnm{{Baxter}},~\bfnm{R.~A.}\binits{R.~A.}} (\byear{1995}).
\btitle{{Finding overlapping distributions with MML.}}
\btype{Technical
Report} No.
\bnumber{{244}},
\binstitution{{Dept.
Computer Science, Monash Univ.}},
\baddress{Clayton, Australia}.
\end{btechreport}
\endbibitem

%
%
\bibitem[\protect\citeauthoryear{{Beal}}{2003}]{Beal03}
\begin{bphdthesis}[author]
\bauthor{\bsnm{{Beal}},~\bfnm{M.~J.}\binits{M.~J.}} (\byear{2003}).
\btitle{{Variational algorithms for approximate Bayesian inference.}}
\btype{Ph.D. thesis}, \bschool{{Gatsby Computational Neuroscience Unit,
Univ. College London}}.
\end{bphdthesis}
\endbibitem

%
%
\bibitem[\protect\citeauthoryear{{Binney} and
{Merrifield}}{1998}]{1998gaasbookB}
\begin{bbook}[author]
\bauthor{\bsnm{{Binney}},~\bfnm{J.}\binits{J.}} \AND
\bauthor{\bsnm{{Merrifield}},~\bfnm{M.}\binits{M.}}
(\byear{1998}). \btitle{{Galactic Astronomy}}. \bpublisher{Princeton
Univ. Press, Princeton, NJ}.
\end{bbook}
\endbibitem

%
%
\bibitem[\protect\citeauthoryear{{Blaauw} et al.}{1960}]{1960MNRAS121123B}
\begin{barticle}[author]
\bauthor{\bsnm{{Blaauw}},~\bfnm{A.}\binits{A.}},
\bauthor{\bsnm{{Gum}},~\bfnm{C.~S.}\binits{C.~S.}},
\bauthor{\bsnm{{Pawsey}},~\bfnm{J.~L.}\binits{J.~L.}} \AND
\bauthor{\bsnm{{Westerhout}},~\bfnm{G.}\binits{G.}}
(\byear{1960}).
\btitle{{The new IAU system of galactic
coordinates (1958 revision)}}.
\bjournal{Mon. Not. R. Astron. Soc.}
\bvolume{121}
\bpages{123}.
\end{barticle}
\endbibitem

%
%
\bibitem[\protect\citeauthoryear{{Bovy}}{2010}]{Bovy10b}
\begin{barticle}[author]
\bauthor{\bsnm{{Bovy}},~\bfnm{J.}\binits{J.}}
(\byear{2010}).
\btitle{{Tracing the Hercules stream around the galaxy}}.
\bjournal{Astrophys. J.} \bvolume{725} \bpages{1676}.
\end{barticle}
\endbibitem

%
%
\bibitem[\protect\citeauthoryear{{Bovy}, {Hogg} and {Roweis}}{2009}]{Bovy09b}
\begin{barticle}[author]
\bauthor{\bsnm{{Bovy}},~\bfnm{J.}\binits{J.}},
\bauthor{\bsnm{{Hogg}},~\bfnm{D.~W.}\binits{D.~W.}} \AND
\bauthor{\bsnm{{Roweis}},~\bfnm{S.~T.}\binits{S.~T.}}
(\byear{2009}). \btitle{{The velocity distribution of nearby stars from
Hipparcos data I. The
significance of the moving groups}}.
\bjournal{Astrophys. J.} \bvolume{700} \bpages{1794}.
\end{barticle}
\endbibitem

%
%
\bibitem[\protect\citeauthoryear{{Bovy} and {Hogg}}{2010}]{Bovy10a}
\begin{barticle}[author]
\bauthor{\bsnm{{Bovy}},~\bfnm{J.}\binits{J.}} \AND
\bauthor{\bsnm{{Hogg}},~\bfnm{D.~W.}\binits{D.~W.}}
(\byear{2010}).
\btitle{{The velocity distribution of nearby stars from
Hipparcos data~II. The
nature of the low-velocity moving groups}}.
\bjournal{Astrophys.~J.}
\bvolume{717}
\bpages{617}.
\end{barticle}
\endbibitem

%
%
\bibitem[\protect\citeauthoryear{{Broniatowski}, {Celeux} and
{Diebolt}}{1983}]{Broniatowksi83}
\begin{bincollection}[author]
\bauthor{\bsnm{{Broniatowski}},~\bfnm{M.}\binits{M.}},
\bauthor{\bsnm{{Celeux}},~\bfnm{G.}\binits{G.}} \AND
\bauthor{\bsnm{{Diebolt}},~\bfnm{J.}\binits{J.}}
(\byear{1983}).
\btitle{{Reconaissance de Densit{\'{e}}s par un
Algorithme d'Apprentissage
Probabiliste}}.
In \bbooktitle{{Data Analysis and Informatics, Vol. 3}}
\bpages{359--373}.
\bpublisher{North-Holland},
\baddress{Amsterdam}.
\end{bincollection}
\MR{0787647}
\endbibitem

%
%
\bibitem[\protect\citeauthoryear{{Carroll} et al.}{2006}]{Carroll06a}
\begin{bbook}[author]
\bauthor{\bsnm{{Carroll}},~\bfnm{R.~J.}\binits{R.~J.}},
\bauthor{\bsnm{{Ruppert}},~\bfnm{D.}\binits{D.}},
\bauthor{\bsnm{{Stefanski}},~\bfnm{L.~A.}\binits{L.~A.}} \AND
\bauthor{\bsnm{{Crainiceanu}},~\bfnm{C.~M.}\binits{C.~M.}}
(\byear{2006}). \btitle{Measurement Error in Nonlinear Models: A Modern
Perspective},
\bedition{2nd} ed.
\bpublisher{{Chapman and Hall/CRC}}, \baddress{Boca Raton, FL}.
\end{bbook}
\MR{2243417}
\endbibitem

%
%
\bibitem[\protect\citeauthoryear{{Celeux} and {Diebolt}}{1985}]{Celeux85a}
\begin{barticle}[author]
\bauthor{\bsnm{{Celeux}},~\bfnm{G.}\binits{G.}} \AND
\bauthor{\bsnm{{Diebolt}},~\bfnm{J.}\binits{J.}}
(\byear{1985}). \btitle{{The {SEM} algorithm: A probabilistic teacher
algorithm derived from
the {EM} algorithm for the mixture problem}}.
\bjournal{{Comput. Statist. Quart}}
\bvolume{2}
\bpages{73}.
\end{barticle}
\endbibitem

%
%
\bibitem[\protect\citeauthoryear{{Celeux} and {Diebolt}}{1986}]{Celeux86b}
\begin{barticle}[author]
\bauthor{\bsnm{{Celeux}},~\bfnm{G.}\binits{G.}} \AND
\bauthor{\bsnm{{Diebolt}},~\bfnm{J.}\binits{J.}}
(\byear{1986}).
\btitle{{L'Algorithme {SEM}: un Algorithme
d'Apprentissage Probabiliste pour la
Reconnaisance de M{\'{e}}langes de Densit{\'{e}}s}}.
\bjournal{{Rev. Stat. Appl.}}
\bvolume{34}
\bpages{35}.
\end{barticle}
\endbibitem

%
%
\bibitem[\protect\citeauthoryear{{De Simone}, {Wu} and
{Tremaine}}{2004}]{2004MNRAS350627D}
\begin{barticle}[author]
\bauthor{\bsnm{{De Simone}},~\bfnm{R.}\binits{R.}},
\bauthor{\bsnm{{Wu}},~\bfnm{X.}\binits{X.}} \AND
\bauthor{\bsnm{{Tremaine}},~\bfnm{S.}\binits{S.}}
(\byear{2004}).
\btitle{{The stellar velocity distribution in the solar
neighbourhood}}.
\bjournal{Mon. Not. R. Astron. Soc.}
\bvolume{350}
\bpages{627}.
\end{barticle}
\endbibitem

%
%
\bibitem[\protect\citeauthoryear{{Dehnen}}{1998}]{1998AJ1152384D}
\begin{barticle}[author]
\bauthor{\bsnm{{Dehnen}},~\bfnm{W.}\binits{W.}} (\byear{1998}).
\btitle{{The distribution of nearby stars in velocity space inferred
from
Hipparcos data}}.
\bjournal{Astron. J.}
\bvolume{115}
\bpages{2384}.
\end{barticle}
\endbibitem

%
%
\bibitem[\protect\citeauthoryear{{Dehnen}}{2000}]{Dehnen00a}
\begin{barticle}[author]
\bauthor{\bsnm{{Dehnen}},~\bfnm{W.}\binits{W.}}
(\byear{2000}).
\btitle{{The effect of the outer Lindblad resonance of the galactic bar
on the
local stellar velocity distribution}}.
\bjournal{Astron. J.}
\bvolume{119}
\bpages{800}.
\end{barticle}
\endbibitem

%
%
\bibitem[\protect\citeauthoryear{{Dehnen} and {Binney}}{1998}]{1998MNRAS298387D}
\begin{barticle}[author]
\bauthor{\bsnm{{Dehnen}},~\bfnm{W.}\binits{W.}}
\AND
\bauthor{\bsnm{{Binney}},~\bfnm{J.~J.}\binits{J.~J.}}
(\byear{1998}).
\btitle{{Local stellar kinematics from Hipparcos
data}}.
\bjournal{Mon. Not. R. Astron. Soc.}
\bvolume{298}
\bpages{387}.
\end{barticle}
\endbibitem

%
%
\bibitem[\protect\citeauthoryear{{Delaigle} and
{Meister}}{2008}]{Delaigle08a}
\begin{barticle}[author]
\bauthor{\bsnm{{Delaigle}},~\bfnm{A.}\binits{A.}} \AND
\bauthor{\bsnm{{Meister}},~\bfnm{A.}\binits{A.}}
(\byear{2008}).
\btitle{Density estimation with heteroscedastic error}.
\bjournal{Bernoulli}
\bvolume{14}
\bpages{562--579}.
\end{barticle}
\MR{2544102}
\endbibitem

%
%
\bibitem[\protect\citeauthoryear{{Dempster}, {Laird} and
{Rubin}}{1977}]{Dempster1977}
\begin{barticle}[author]
\bauthor{\bsnm{{Dempster}},~\bfnm{A.~P.}\binits{A.~P.}},
\bauthor{\bsnm{{Laird}},~\bfnm{N.~M.}\binits{N.~M.}} \AND
\bauthor{\bsnm{{Rubin}},~\bfnm{D.~B}\binits{D.~B.}}
(\byear{1977}).
\btitle{{Maximum likelihood from incomplete data via
the EM algorithm}}.
\bjournal{J. R. Stat. Soc.
Ser. B Methodol. Stat.}
\bvolume{39}
\bpages{1--38}.
\end{barticle}
\MR{0501537}
\endbibitem

%
%
\bibitem[\protect\citeauthoryear{{Diebolt} and {Robert}}{1994}]{diebolt94a}
\begin{barticle}[author]
\bauthor{\bsnm{{Diebolt}},~\bfnm{J.}\binits{J.}} \AND
\bauthor{\bsnm{{Robert}},~\bfnm{C.~P.}\binits{C.~P.}}
(\byear{1994}).
\btitle{{Estimation of finite mixture distributions
through Bayesian
sampling}}.
\bjournal{J. R. Stat. Soc.
Ser. B Methodol. Stat.}
\bvolume{56}
\bpages{363--375}.
\end{barticle}
\MR{1281940}
\endbibitem

%
%
\bibitem[\protect\citeauthoryear{{ESA}}{1997}]{ESA97a}
\begin{bbook}[author]
\bauthor{\bsnm{{ESA}}~}
(\byear{1997}).
\btitle{{The \textit{Hipparcos}
and Tycho Catalogues}}.
\bpublisher{{ESA SP-1200}},
\baddress{Noordwijk}.
\end{bbook}
\endbibitem

%
%
\bibitem[\protect\citeauthoryear{{Famaey} et al.}{2005}]{2005AA430165F}
\begin{barticle}[author]
\bauthor{\bsnm{{Famaey}},~\bfnm{B.}\binits{B.}},
\bauthor{\bsnm{{Jorissen}},~\bfnm{A.}\binits{A.}},
\bauthor{\bsnm{{Luri}},~\bfnm{X.}\binits{X.}},
\bauthor{\bsnm{{Mayor}},~\bfnm{M.}\binits{M.}},
\bauthor{\bsnm{{Udry}},~\bfnm{S.}\binits{S.}},
\bauthor{\bsnm{{Dejonghe}},~\bfnm{H.}\binits{H.}} \AND
\bauthor{\bsnm{{Turon}},~\bfnm{C.}\binits{C.}}
(\byear{2005}).
\btitle{{Local kinematics of $K$ and $M$ giants from
CORAVEL/Hipparcos/\break Tycho-2
data. Revisiting the concept of superclusters}}.
\bjournal{Astron.  Astrophys.}
\bvolume{430}
\bpages{165}.
\end{barticle}
\endbibitem

%
%
\bibitem[\protect\citeauthoryear{{Gelman} et al.}{2000}]{Gelman00a}
\begin{bbook}[author]
\bauthor{\bsnm{{Gelman}},~\bfnm{A.}\binits{A.}},
\bauthor{\bsnm{{Carlin}},~\bfnm{J.~B.}\binits{J.~B.}},
\bauthor{\bsnm{{Stern}},~\bfnm{H.~S.}\binits{H.~S.}} \AND
\bauthor{\bsnm{{Rubin}},~\bfnm{D.~B.}\binits{D.~B.}}
(\byear{2000}).
\btitle{{Bayesian Data Analysis}}.
\bpublisher{{Chapman
and Hall/CRC, Boca Raton, FL}}.
\end{bbook}
\MR{1385925}
\endbibitem

%
%
\bibitem[\protect\citeauthoryear{{Ghahramani} and
{Beal}}{2000}]{Ghahramani00variationalinference}
\begin{binproceedings}[author]
\bauthor{\bsnm{{Ghahramani}},~\bfnm{Z.}\binits{Z.}} \AND
\bauthor{\bsnm{{Beal}},~\bfnm{M.~J.}\binits{M.~J.}}
(\byear{2000}).
\btitle{{Variational inference for Bayesian mixtures of
factor analysers}}.
In \bbooktitle{Advances in Neural Information
Processing Systems \textit{12}}
(\beditor{\bfnm{S.~A.}\binits{S.~A.}~\bsnm{{Solla}}},
\beditor{\bfnm{T.~K.}\binits{T.~K.}~\bsnm{{Leen}}} \AND
\beditor{\bfnm{K.~R.}\binits{K.~R.}~\bsnm{{Muller}}}, eds.)
\bpages{449}.
\bpublisher{MIT Press, Cambridge, MA}.
\end{binproceedings}
\endbibitem

%
%
\bibitem[\protect\citeauthoryear{{Ghahramani} and
{Jordan}}{1994a}]{Ghahramani1994b}
\begin{btechreport}[author]
\bauthor{\bsnm{{Ghahramani}},~\bfnm{Z.}\binits{Z.}} \AND
\bauthor{\bsnm{{Jordan}},~\bfnm{M.~I.}\binits{M.~I.}}
(\byear{1994}a).
\btitle{{Learning from incomplete data}}.
\binstitution{{CBCL Technical Report No.
108. Center for Biological and
Computational Learning, MIT}}.
\end{btechreport}
\endbibitem

%
%
\bibitem[\protect\citeauthoryear{{Ghahramani} and
{Jordan}}{1994b}]{Ghahramani1994}
\begin{binproceedings}[author]
\bauthor{\bsnm{{Ghahramani}},~\bfnm{Z.}\binits{Z.}} \AND
\bauthor{\bsnm{{Jordan}},~\bfnm{M.~I.}\binits{M.~I.}}
(\byear{1994}b).
\btitle{{Supervised learning from incomplete data via
an EM approach}}.
In \bbooktitle{{Advances in Neural Information
Processing Systems \textit{6}}}
(\beditor{\bfnm{J.~D.}\binits{J.~D.}~\bsnm{{Cowan}}},
\beditor{\bfnm{G.}\binits{G.}~\bsnm{{Tesauro}}} \AND
\beditor{\bfnm{J.}\binits{J.}~\bsnm{{Alspector}}}, eds.)
\bpages{120--127}.
\bpublisher{{Morgan Kaufman}},
\baddress{San Francisco}.
\end{binproceedings}
\endbibitem

%
%
\bibitem[\protect\citeauthoryear{{Helmi} et al.}{1999}]{1999Natur40253H}
\begin{barticle}[author]
\bauthor{\bsnm{{Helmi}},~\bfnm{A.}\binits{A.}},
\bauthor{\bsnm{{White}},~\bfnm{S.~D.~M.}\binits{S.~D.~M.}},
\bauthor{\bsnm{{de Zeeuw}},~\bfnm{P.~T.}\binits{P.~T.}} \AND
\bauthor{\bsnm{{Zhao}},~\bfnm{H.}\binits{H.}}
(\byear{1999}).
\btitle{{Debris streams in the solar neighbourhood as
relicts from the
formation of the milky way}}.
\bjournal{Nature}
\bvolume{402}
\bpages{53--55}.
\end{barticle}
\endbibitem

%
%
\bibitem[\protect\citeauthoryear{{Hogg} et al.}{2005}]{2005ApJ629268H}
\begin{barticle}[author]
\bauthor{\bsnm{{Hogg}},~\bfnm{D.~W.}\binits{D.~W.}},
\bauthor{\bsnm{{Blanton}},~\bfnm{M.~R.}\binits{M.~R.}},
\bauthor{\bsnm{{Roweis}},~\bfnm{S.~T.}\binits{S.~T.}} \AND
\bauthor{\bsnm{{Johnston}},~\bfnm{K.~V.}\binits{K.~V.}}
(\byear{2005}).
\btitle{{Modeling complete distributions with
incomplete observations: The
velocity ellipsoid from Hipparcos data}}.
\bjournal{Astrophys.~J.}
\bvolume{629}
\bpages{268}.
\end{barticle}
\endbibitem

%
%
\bibitem[\protect\citeauthoryear{{Holmberg}, {Nordstr{\"{o}}m} and
{Andersen}}{2009}]{2008arXiv08113982H}
\begin{barticle}[author]
\bauthor{\bsnm{{Holmberg}},~\bfnm{J.}\binits{J.}},
\bauthor{\bsnm{{Nordstr{\"{o}}m}},~\bfnm{B.}\binits{B.}} \AND
\bauthor{\bsnm{{Andersen}},~\bfnm{J.}\binits{J.}}
(\byear{2009}).
\btitle{{The Geneva--Copenhagen survey of the solar
neighbourhood III. Improved
distances, ages, and kinematics}}.
\bjournal{Astron.  Astrophys.}
\bvolume{501}
\bpages{941}.
\end{barticle}
\endbibitem

%
%
\bibitem[\protect\citeauthoryear{{Jasra}, {Holmes} and
{Stephens}}{2005}]{Jasra05a}
\begin{barticle}[author]
\bauthor{\bsnm{{Jasra}},~\bfnm{A.}\binits{A.}},
\bauthor{\bsnm{{Holmes}},~\bfnm{C.~C.}\binits{C.~C.}} \AND
\bauthor{\bsnm{{Stephens}},~\bfnm{D.~A.}\binits{D.~A.}}
(\byear{2005}).
\btitle{{{M}arkov {c}hain {M}onte {C}arlo methods and
the label switching
problem in {B}ayesian mixture modeling}}.
\bjournal{Statist. Sci.}
\bvolume{20}
\bpages{50--67}.
\end{barticle}
\MR{2182987}
\endbibitem

%
%
\bibitem[\protect\citeauthoryear{{MacKay}}{2003}]{Mackay2003}
\begin{bbook}[author]
\bauthor{\bsnm{{MacKay}},~\bfnm{D.~J.~C.}\binits{D.~J.~C.}}
(\byear{2003}).
\btitle{{Information Theory, Inference, and Learning
Algorithms}}.
\bpublisher{Cambridge Univ. Press},
\baddress{Cambridge}.
\end{bbook}
\MR{2012999}
\endbibitem

%
%
\bibitem[\protect\citeauthoryear{{McLachlan} and
{Basford}}{1988}]{McLachlan1988}
\begin{bbook}[author]
\bauthor{\bsnm{{McLachlan}},~\bfnm{G.~J.}\binits{G.~J.}} \AND
\bauthor{\bsnm{{Basford}},~\bfnm{K.}\binits{K.}}
(\byear{1988}).
\btitle{{Mixture Models: Inference and Application to
Clustering}}.
\bpublisher{Dekker},
\baddress{New York}.
\end{bbook}
\MR{0926484}
\endbibitem

%
%
\bibitem[\protect\citeauthoryear{{Nordstr{\"{o}}m} et al.}{2004}]{2004AA418989N}
\begin{barticle}[author]
\bauthor{\bsnm{{Nordstr{\"{o}}m}},~\bfnm{B.}\binits{B.}},
\bauthor{\bsnm{{Mayor}},~\bfnm{M.}\binits{M.}},
\bauthor{\bsnm{{Andersen}},~\bfnm{J.}\binits{J.}},
\bauthor{\bsnm{{Holmberg}},~\bfnm{J.}\binits{J.}},
\bauthor{\bsnm{{Pont}},~\bfnm{F.}\binits{F.}},
\bauthor{\bsnm{{J{\o}rgen\-sen}},~\bfnm{B.~R.}\binits{B.~R.}},
\bauthor{\bsnm{{Olsen}},~\bfnm{E.~H.}\binits{E.~H.}},
\bauthor{\bsnm{{Udry}},~\bfnm{S.}\binits{S.}} \AND
\bauthor{\bsnm{{Mowlavi}},~\bfnm{N.}\binits{N.}}
(\byear{2004}).
\btitle{{The Geneva--Copenha\-gen survey of the solar
neighbourhood. Ages,
metallicities, and kinematic properties of {$\sim\!$}14 000 $F$ and $G$ dwarfs}}.
\bjournal{Astron.  Astrophys.}
\bvolume{418}
\bpages{989}.
\end{barticle}
\endbibitem

%
%
\bibitem[\protect\citeauthoryear{{Oliver}, {Baxter} and
{Wallace}}{1996}]{Oliver96a}
\begin{binproceedings}[author]
\bauthor{\bsnm{{Oliver}},~\bfnm{J.~J.}\binits{J.~J.}},
\bauthor{\bsnm{{Baxter}},~\bfnm{R.~A.}\binits{R.~A.}} \AND
\bauthor{\bsnm{{Wallace}},~\bfnm{C.~S.}\binits{C.~S.}}
(\byear{1996}).
\btitle{{Unsupervised learning using MML}}. In
\bbooktitle{Machine Learning: Proceedings of the Thirteenth
International
Conference (ICML 96)}
\bpages{364}.
\bpublisher{Morgan Kaufmann},
\baddress{San Francisco}.
\end{binproceedings}
\endbibitem

%
%
\bibitem[\protect\citeauthoryear{{Ormoneit} and {Tresp}}{1996}]{Ormoneit1995}
\begin{binproceedings}[author]
\bauthor{\bsnm{{Ormoneit}},~\bfnm{D.}\binits{D.}} \AND
\bauthor{\bsnm{{Tresp}},~\bfnm{V.}\binits{V.}}
(\byear{1996}).
\btitle{{Improved Gaussian mixture density estimates
using Bayesian penalty
terms and network averaging}}.
In \bbooktitle{{Advances in Neural Information Processing Systems \textit{8},
NIPS,
Denver, CO, November 27--30, 1995}}
(\beditor{\bfnm{D.~S.}\binits{D.~S.}~\bsnm{{Touretzky}}},
\beditor{\bfnm{M.}\binits{M.}~\bsnm{{Mozer}}} \AND
\beditor{\bfnm{M.~E.}\binits{M.~E.}~\bsnm{{Hasselmo}}}, eds.)
\bpages{542--548}.
\bpublisher{MIT Press},
\baddress{Cambridge}.
\end{binproceedings}
\endbibitem

%
%
\bibitem[\protect\citeauthoryear{{Quillen} and
{Minchev}}{2005}]{2005AJ130576Q}
\begin{barticle}[author]
\bauthor{\bsnm{{Quillen}},~\bfnm{A.~C.}\binits{A.~C.}} \AND
\bauthor{\bsnm{{Minchev}},~\bfnm{I.}\binits{I.}}
(\byear{2005}).
\btitle{{The effect of spiral structure on the stellar
velocity distribution in
the solar neighborhood}}.
\bjournal{Astron. J.}
\bvolume{130}
\bpages{576}.
\end{barticle}
\endbibitem

%
%
\bibitem[\protect\citeauthoryear{{Rabiner} and
{Biing-Hwang}}{1993}]{Rabiner1993}
\begin{bbook}[author]
\bauthor{\bsnm{{Rabiner}},~\bfnm{L.}\binits{L.}} \AND
\bauthor{\bsnm{{Biing-Hwang}},~\bfnm{J.}\binits{J.}}
(\byear{1993}).
\btitle{{Fundamentals of Speech Recognition}}.
\bpublisher{Prentice-Hall},
\baddress{New York}.
\end{bbook}
\endbibitem

%
%
\bibitem[\protect\citeauthoryear{Rasmussen}{2000}]{rasmussen00infinite}
\begin{binproceedings}[author]
\bauthor{\bsnm{Rasmussen},~\bfnm{C.}\binits{C.}} (\byear{2000}).
\btitle{{The infinite Gaussian mixture model}}. In
\bbooktitle{{Advances in Neural Information Processing Systems \textit{12}}}
(\beditor{\bfnm{S.~A.}\binits{S.~A.}~\bsnm{{Solla}}},
\beditor{\bfnm{T.~K.}\binits{T.~K.}~\bsnm{{Leen}}} \AND
\beditor{\bfnm{K.~R.}\binits{K.~R.}~\bsnm{{Muller}}}, eds.)
\bpages{554--560}.
\bpublisher{MIT Press},
\baddress{Cambridge}.
\end{binproceedings}
\endbibitem

%
%
\bibitem[\protect\citeauthoryear{{Richardson} and
{Green}}{1997}]{Richardson97a}
\begin{barticle}[author]
\bauthor{\bsnm{{Richardson}},~\bfnm{S.}\binits{S.}} \AND
\bauthor{\bsnm{{Green}},~\bfnm{P.~J.}\binits{P.~J.}}
(\byear{1997}).
\btitle{{On Bayesian analysis of mixtures with an
unknown number of
components}}.
\bjournal{J. R. Stat. Soc. Ser. B
Methodol. Stat.}
\bvolume{59}
\bpages{731--792}.
\end{barticle}
\MR{1483213}
\endbibitem

%
%
\bibitem[\protect\citeauthoryear{{Rissanen}}{1978}]{rissanen78a}
\begin{barticle}[author]
\bauthor{\bsnm{{Rissanen}},~\bfnm{J.}\binits{J.}}
(\byear{1978}).
\btitle{{Modeling by shortest data description}}.
\bjournal{{Automatica}}
\bvolume{14}
\bpages{465}.
\end{barticle}
\endbibitem

%
%
\bibitem[\protect\citeauthoryear{{Roberts} et~al.}{1998}]{roberts98a}
\begin{barticle}[author]
\bauthor{\bsnm{{Roberts}},~\bfnm{S.~J.}\binits{S.~J.}},
\bauthor{\bsnm{{Husmeier}},~\bfnm{D.}\binits{D.}},
\bauthor{\bsnm{{Rezek}},~\bfnm{I.}\binits{I.}} \AND
\bauthor{\bsnm{{Penny}},~\bfnm{W.}\binits{W.}}
(\byear{1998}).
\btitle{{Bayesian approaches to Gaussian mixture
modeling}}.
\bjournal{IEEE Trans. Pattern Anal. Mach.
Intell.}
\bvolume{20}
\bpages{1133}.
\end{barticle}
\endbibitem

%
%
\bibitem[\protect\citeauthoryear{{Schafer}}{1993}]{Schafer93a}
\begin{barticle}[author]
\bauthor{\bsnm{{Schafer}},~\bfnm{D.~W.}\binits{D.~W.}} (\byear{1993}).
\btitle{Likelihood analysis for probit regression with measurement
errors}.
\bjournal{Biometrika}
\bvolume{80}
\bpages{899}.
\end{barticle}
\endbibitem

%
%
\bibitem[\protect\citeauthoryear{{Schafer} and {Purdy}}{1996}]{Schafer96a}
\begin{barticle}[author]
\bauthor{\bsnm{{Schafer}},~\bfnm{D.~W.}\binits{D.~W.}} \AND
\bauthor{\bsnm{{Purdy}},~\bfnm{K.~G.}\binits{K.~G.}}
(\byear{1996}).
\btitle{Likelihood analysis for errors-in-variables
regression with replicate
measurements}.
\bjournal{Biometrika}
\bvolume{83}
\bpages{813--824}.
\end{barticle}
\MR{1440046}
\endbibitem

%
%
\bibitem[\protect\citeauthoryear{{Schwartz}}{1978}]{Schwartz78a}
\begin{barticle}[author]
\bauthor{\bsnm{{Schwartz}},~\bfnm{G.}\binits{G.}}
(\byear{1978}).
\btitle{{Estimating the dimension of a model}}.
\bjournal{{Ann. Statist.}}
\bvolume{{6}}
\bpages{461--464}.
\end{barticle}
\MR{0468014}
\endbibitem

%
%
\bibitem[\protect\citeauthoryear{{Silverman}}{1986}]{Silverman86a}
\begin{bbook}[author]
\bauthor{\bsnm{{Silverman}},~\bfnm{B.~W.}\binits{B.~W.}}
(\byear{1986}). \btitle{{Density Estimation for Statistics and Data
Analysis}}. \bpublisher{{Chapman and Hall, Boca Raton, FL}}.
\end{bbook}
\MR{0848134}
\endbibitem

%
%
\bibitem[\protect\citeauthoryear{{Skuljan}, {Hearnshaw} and
{Cottrell}}{1999}]{1999MNRAS308731S}
\begin{barticle}[author]
\bauthor{\bsnm{{Skuljan}},~\bfnm{J.}\binits{J.}},
\bauthor{\bsnm{{Hearnshaw}},~\bfnm{J.~B.}\binits{J.~B.}} \AND
\bauthor{\bsnm{{Cottrell}},~\bfnm{P.~L.}\binits{P.~L.}}
(\byear{1999}).
\btitle{{Velocity distribution of stars in the solar
neighbourhood}}.
\bjournal{Mon. Not. R. Astron. Soc.}
\bvolume{308}
\bpages{731}.
\end{barticle}
\endbibitem

%
%
\bibitem[\protect\citeauthoryear{{Staudenmayer}, {Ruppert} and
{Buonaccorsi}}{2008}]{Staudenmayer08a}
\begin{barticle}[author]
\bauthor{\bsnm{{Staudenmayer}},~\bfnm{J.}\binits{J.}},
\bauthor{\bsnm{{Ruppert}},~\bfnm{D.}\binits{D.}} \AND
\bauthor{\bsnm{{Buonaccorsi}},~\bfnm{J.}\binits{J.}}
(\byear{2008}).
\btitle{Density estimation in the presence of
heteroscedastic measurement
error}.
\bjournal{J. Amer. Statist. Assoc.}
\bvolume{103}
\bpages{726--736}.
\end{barticle}
\MR{2524005}
\endbibitem

%
%
\bibitem[\protect\citeauthoryear{{Stefanski} and
{Carroll}}{1990}]{Stefanski90a}
\begin{barticle}[author]
\bauthor{\bsnm{{Stefanski}},~\bfnm{L.~A.}\binits{L.~A.}} \AND
\bauthor{\bsnm{{Carroll}},~\bfnm{R.~J.}\binits{R.~J.}}
(\byear{1990}).
\btitle{{Deconvoluting kernel density estimators}}.
\bjournal{Statistics}
\bvolume{21}
\bpages{169--184}.
\end{barticle}
\MR{1054861}
\endbibitem

%
%
\bibitem[\protect\citeauthoryear{{Stone}}{1974}]{stone74a}
\begin{barticle}[author]
\bauthor{\bsnm{{Stone}},~\bfnm{M.}\binits{M.}}
(\byear{1974}).
\btitle{{Cross-validation choice and assessment of statistical
predictions}}.
\bjournal{J. R. Stat. Soc.
Ser. B Methodol. Stat.}
\bvolume{36}
\bpages{111--147}.
\end{barticle}
\MR{0356377}
\endbibitem

%
%
\bibitem[\protect\citeauthoryear{{Ueda} et al.}{1998}]{Naonori1998}
\begin{binproceedings}[author]
\bauthor{\bsnm{{Ueda}},~\bfnm{N.}\binits{N.}},
\bauthor{\bsnm{{Nakano}},~\bfnm{R.}\binits{R.}},
\bauthor{\bsnm{{Ghahramani}},~\bfnm{Z.}\binits{Z.}} \AND
\bauthor{\bsnm{{Hinton}},~\bfnm{G.~E.}\binits{G.~E.}}
(\byear{1998}).
\btitle{{Split and merge EM algorithm for improving
Gaussian mixture density
estimates}}.
In \bbooktitle{{Neural Networks for Signal Processing VIII, 1998.
Proceedings
of the 1998 IEEE Signal Processing Society Workshop}}
\bpages{274--283}. \bpublisher{IEEE}.
\end{binproceedings}
\endbibitem

%
%
\bibitem[\protect\citeauthoryear{{van Leeuwen}}{2007a}]{2007ASSL250V}
\begin{bbook}[author]
\bauthor{\bsnm{{van Leeuwen}},~\bfnm{F.}\binits{F.}}
(\byear{2007}a).
\btitle{{Hipparcos, the New Reduction of the Raw Data}}.
\bseries{Astrophysics and Space Science Library}
\bvolume{250}.
\bpublisher{Springer},
\baddress{Dordrecht}.
\end{bbook}
\endbibitem

%
%
\bibitem[\protect\citeauthoryear{{van Leeuwen}}{2007b}]{2007AA474653V}
\begin{barticle}[author]
\bauthor{\bsnm{{van Leeuwen}},~\bfnm{F.}\binits{F.}}
(\byear{2007}b).
\btitle{{Validation of the new Hipparcos reduction}}.
\bjournal{Astron. Astrophys.}
\bvolume{474}
\bpages{653}.
\end{barticle}
\endbibitem

%
%
\bibitem[\protect\citeauthoryear{{Wallace} and {Boulton}}{1968}]{wallace68a}
\begin{barticle}[author]
\bauthor{\bsnm{{Wallace}},~\bfnm{C.~S.}\binits{C.~S.}} \AND
\bauthor{\bsnm{{Boulton}},~\bfnm{D.~M.}\binits{D.~M.}}
(\byear{1968}).
\btitle{{An information measure for classification}}.
\bjournal{{Comput.~J.}}
\bvolume{11}
\bpages{185}.
\end{barticle}
\endbibitem

%
%
\bibitem[\protect\citeauthoryear{{Wu}}{1983}]{Wu1983}
\begin{barticle}[author]
\bauthor{\bsnm{{Wu}},~\bfnm{C.~F.~J}\binits{C.~F.~J.}}
(\byear{1983}).
\btitle{{On the convergence properties of the EM algorithm}}.
\bjournal{Ann. Statist.}
\bvolume{11}
\bpages{95--103}.
\end{barticle}
\MR{0684867}
\endbibitem

%
%
\bibitem[\protect\citeauthoryear{{Zhang}}{1990}]{Zhang90a}
\begin{barticle}[author]
\bauthor{\bsnm{{Zhang}},~\bfnm{C.~H.}\binits{C.~H.}}
(\byear{1990}).
\btitle{Fourier methods for estimating mixing densities and
distributions}.
\bjournal{Ann. Statist.}
\bvolume{18}
\bpages{806--831}.
\end{barticle}
\MR{1056338}
\endbibitem

\end{thebibliography}
\end{document}